\documentclass[aps,pre,twocolumn,floatfix,superscriptaddress,amsfonts,amssymb,a4paper]{revtex4}
\usepackage[T1]{fontenc}
\usepackage[utf8]{inputenc}
\usepackage{amsmath}
\usepackage{amscd}
\usepackage{graphicx}
\usepackage{afterpage}

\newcommand{\kb}{\overline{k}}
\newcommand{\tn}[2]{n_{#1,#2}}

\newcommand{\av}[1]{\left\langle #1\right\rangle}

\newcommand{\pf}{p^{(f)}}
\newcommand{\pg}{p^{(g)}}

\newcommand{\zf}{z^{(f)}}
\newcommand{\zg}{z^{(g)}}

\newlength{\figheight}
\begin{document}
\title{Correlations in connected random graphs}
\author{Piotr Bialas}
\email[email: ]{pbialas@th.if.uj.edu.pl}\affiliation{Marian Smoluchowski Institute of Physics, Jagellonian University, Reymonta 4, 30--059 Krakow, Poland}
\affiliation{Mark Kac Complex Systems Research Centre, Faculty of Physics, Astronomy and Applied Computer Science, \\
Jagellonian University, Reymonta 4, 30--059 Krakow, Poland}
\author{Andrzej K. Oleś}
\email[email: ]{oles@th.if.uj.edu.pl}\affiliation{Marian Smoluchowski Institute of Physics, Jagellonian University, Reymonta 4, 30--059 Krakow, Poland}

\begin{abstract}
We study the properties of the giant connected component in random
graphs with arbitrary degree distribution. We concentrate on the
degree-degree correlations. We show that the adjoining nodes in the
giant connected component are correlated and derive analytic
formulas for the joint nearest-neighbor degree probability
distribution. Using those results we describe correlations in
maximal entropy connected random graphs. We show that connected
graphs are disassortative and that correlations are strongly related
to the presence of one-degree nodes (leaves). We propose an
efficient algorithm for generating connected random graphs. We
illustrate our results with several examples.
\vspace{0mm}\\\\
\begin{tabular}{p{7.3cm} p{6.6cm}}
Published in: Phys. Rev. E {\bf 77}, 036124 (2008). &PACS numbers: 89.75.Hc, 05.10.--a, 05.90.+m
\end{tabular}
\end{abstract}
\maketitle

\section{Introduction}

In the last decade or so, there has been a great increase of
interest in the theory of random graphs and networks (in the
following we will use those two terms interchangeably). While in
principle this is a branch of mathematics, much of this effort was
fueled by the availability of ``experimental'' data on real graphs
(see \cite{review} for review).  These data are compared to the
predictions of various random graphs models.  Probably the best
known and simplest example of such reference models is the ensemble
of all labeled graphs with $V$ vertices and $L$ links (without
multiple- and self-links), chosen with uniform probability.  We
will call this model Erd\"{o}s-R\'{e}nyi (ER) graphs after the
authors, who were the first to introduce and study them
\cite{ErdosRenyi}.

The ER ensemble is the simplest example of the so-called ``maximally
random'' graphs. Intuitively those are the ensembles where the
distributions of vertices and links joining them are ``as random as
possible'' for a given set of constraints. In the case of ER graphs
the only constraints are the fixed number of links and vertices. The
``maximal randomness'' can be formalized using the notion of {\em
  entropy} (see next section). The maximally random ensembles serve as
null hypothesis. For example, it was the deviation of data collected
on the World Wide Web (WWW) graph from the predictions of the ER
model that triggered the interest in random networks \cite{nature},
because it implied that those graphs were not created just by
joining vertices at random, but required the existence of another
mechanism \cite{preferential}.

A popular generalization of the ER ensemble are graphs with a given
degree distribution (degree of a node is the number of links
attached to it)
\cite{MolloyReed,Newman2001,BCK,BauerBernard,BurdaKrzywicki,fronczak}.
One feature of those ensembles is the absence of correlations
between neighboring nodes' degrees, at least for degree
distributions without heavy tails (see the discussion in Sec.~\ref{sec:sf}). The object of our study was to
find what happens when we constrain to connected graphs only.  A
simple argument indicated that correlations would appear: a neighbor
of a node with degree one (leaf) must have its degree greater than
1; otherwise, they would form a separate connected component.
Similarly, all neighbors of a node cannot have their degree equal
to 1, as such a ``hedgehog'' would also form a separate connected
component \cite{pb1,oles}. This obviously leads to correlations. It
is not clear, however, how strong they are and if they survive the
large-$V$ limit.  We have already studied those correlations
numerically in Ref.~\cite{oles} and found that they also appear in
large graphs. In this paper we derive the analytic formulas
describing them. We also found a strong indication that the
described mechanism is the only one responsible for the correlation
in maximally random connected graphs: when we forbid vertices with
degree 1 correlations disappear.

Connectivity is a nonlocal constraint hard to deal with.  To study
the properties of connected graphs we use another feature of
maximally random graphs with a given degree distribution: the
appearance of a connected component that includes a finite fraction
of all the vertices (and links). From the properties of this giant
connected component we can infer the properties of connected graphs.

The paper is organized as follows: Section~\ref{sec:basic}
introduces some basic definitions concerning random graphs.  In
Sec.~\ref{sec:connected} we present the method of generating
functions used to study the properties of the giant connected
component in random graphs with arbitrary degree distribution
\cite{Newman2001}. Then we calculate degree-degree correlations in
the giant component. Section~\ref{sec:examples}  contains some
examples where we compare our predictions with the results of Monte
Carlo (MC) simulations. Finally, we show in
Sec.~\ref{sec:connectedgraphs} how to relate connected random graphs
to giant connected components in other ensembles. In
Sec.~\ref{sec:uncorrelatedconnectedgraphs} we address the situation
when correlations in random graphs are suppressed by the absence of
vertices with degree one (leaves). The paper is summarized in
Sec.~\ref{sec:summary}.

\section{Random  Graphs}
\label{sec:basic}

\subsection{Average degree}

Formally we consider random graphs as an ensemble of graphs
$\mathcal{G}$ with probability $P(G)$ assigned to every graph
$G\in\mathcal{G}$. Using this definition we introduce the entropy
of the ensemble:
\begin{equation}\label{def:entropy}
S=-\sum_{G\in\mathcal{G}}P(G)\ln P(G).
\end{equation}
The maximally random ensembles described in the previous section are
those which for given constraints have maximal entropy.

Denoting by $O(G)$ some property of graph $G$ we can calculate its
average over the whole ensemble:
\begin{equation}\label{eq:avo}
\av{O}_G=\sum_{G\in\mathcal{G}} O(G) P(G).
\end{equation}
The most widely studied example is the probability distribution of
node degrees:
\begin{equation}\label{eq:avpq}
p_k=\av{\frac{n_k(G)}{V(G)}}_G,
\end{equation}
where $n_k(G)$ is the number of vertices with degree $k$ and $V(G)$
is the total number of vertices in graph $G$ (in the following we
will often omit the argument $G$).  The mean of this distribution is
the ``link density,''
\begin{equation}\label{eq:z}
\av{k}=\sum_k k p_k = \av{\frac{2 L(G)}{V(G)}}_G\equiv z,
\end{equation}
because $\sum_k k n_k = 2 L(G)$; by $L(G)$, we denote the number of
links in graph $G$.

However, what is frequently observed  is not an average
\eqref{eq:avo}, but the properties of a single graph (e.g., WWW).
That is why we are actually interested in the probability that our
model will produce a graph with those properties. It is described by
the distribution
\begin{equation}\label{eq:disto}
P(O)=\sum_{G\in\mathcal{G}} \delta(O-O(G)) P(G).
\end{equation}
In many cases this distribution is sufficiently well characterized
by its mean \eqref{eq:avo} with relative fluctuations disappearing
in the large-$V$ limit. In this situation we will say the $O$ is
self-averaging. In such a case one can infer the properties of the
whole ensemble from the properties of just one large graph. We want
 to emphasize, however, that this is only an assumption that has to be
checked for each particular model (see \cite{Santo} for a discussion
of self-averaging in real graphs).

In Appendix~\ref{app:non-sa} we show for illustration a definition
of a non-self-averaging ensemble.  Although this is an artificial
example, let it serve as a warning. In this paper we assume that our
models are self-averaging without any further formal proofs.

We end with the following comment: as in the self-averaging
ensemble fluctuations do not matter, in the large-volume limit we
have
\begin{equation}
p_k=\av{\frac{n_k(G)}{V(G)}}_G \sim \frac{\av{n_k(G)}_G}{\av{V(G)}_G}.
\end{equation}
We will use this kind of approximations in the following sections.

\subsection{Correlations}

The distribution $p_k$ does not give any information about the
correlations between vertices. An obvious generalization is the
joint distribution $p_{q,r}$ which describes the probability that a
pair of nearest neighbors (NNs) has degrees $q$ and $r$ (we assume
that we pick a pair of NNs with uniform probability):
\begin{eqnarray}\label{eq:avpqr}
p_{q,r}=\av{\frac{\tn{q}{r}}{2L}},
\end{eqnarray}
where $\tn{q}{r}$ is the number of links with their start point
having degree $q$ and endpoint having degree $r$.  Note that we
treat each undirected link as two directed links. On an undirected
graph,
\begin{equation}
\tn qr=\tn rq,\quad \sum_{q,r}\tn qr = 2L\quad \text{and}\quad \sum_{r}\tn qr = q n_q .
\end{equation}
If vertex degrees are independent, the probability \eqref{eq:avpqr}
should factorize:
\begin{equation}\label{eq:indepqr}
p_{q,r}=\widetilde{p}_q \widetilde{p}_r,\quad \widetilde{p}_q =
\sum_r p_{q,r},
\end{equation}
leading to the relation
\begin{equation}\label{eq:indepnqr}
\av{\frac{\tn qr}{2L}} =q r\av{\frac{n_q}{2L}}\av{\frac{n_r}{2L}}.
\end{equation}
One should, however, keep in mind that this defines the absence of
correlations in the {\em ensemble} of graphs. A more appropriate
question could be, are the vertices on individual graphs
uncorrelated (see previous section)? The condition for absence of
correlations between vertices in each individual graph $G$ is
\begin{equation}\label{eq:indepnqr-s}
\frac{\tn qr(G)}{2L(G)} =q r\frac{n_q(G)}{2L(G)}
\frac{n_r(G)}{2L(G)}
\end{equation}
or, after averaging,
\begin{equation}\label{eq:indepavnqr-s}
\av{\frac{\tn qr}{2L}} =q r\av{\frac{n_q}{2L} \frac{n_r}{2L}}.
\end{equation}
As already pointed out, for a large class of ensembles conditions
\eqref{eq:indepnqr} and \eqref{eq:indepavnqr-s} are equivalent in
the large-volume limit. However, it is easy to check that for the
non-self-averaging ensemble in Appendix~\ref{app:non-sa} vertices
on each individual graph are uncorrelated according to the condition
\eqref{eq:indepavnqr-s}, but correlated according to
\eqref{eq:indepnqr}.  Again, we leave this as a warning and proceed
further with the assumption that our models are self-averaging and
that those two conditions are equivalent.

In practice, checking the condition \eqref{eq:indepqr} is difficult
as it entails measuring a two dimensional distribution with good
accuracy. Therefore we introduce another quantity \cite{Satorras2001}
\begin{equation}\label{eq:qb}
\kb(k)=\sum_q \av{\frac{ q \tn{k}{q}}{k n_k}}.
\end{equation}
It describes the average degree of nearest neighbors of a vertex
with degree $k$.  Obviously $\kb(k)$ is defined for a given $k$ only
if $n_k\!>\!0$.  $\kb(k)$ can be interpreted as the first moment of the
conditional probability:
\begin{equation}
p(q|k)=\frac{p_{q,k}}{\widetilde{p}_k}.
\end{equation}
Assuming self-averaging,
\begin{equation}
\kb(k)\approx \sum_q q p(q|k).
\end{equation}
If the degrees are independent, $\kb(k)$ should not depend on $k$ and
\eqref{eq:indepavnqr-s} implies
\begin{equation}\label{eq:indepqbq}
\kb(k)=\sum_q q^2 \av{\frac{ n_q}{2L}}\approx
\frac{\av{k^2}}{\av{k}}.
\end{equation}
When $\kb(k)$ grows with $k$  the graph is called {\em assortative}
and when it shrinks {\em disassortative}.

\section{Connected components}
\label{sec:connected}

In general, maximally random graphs with a given degree distribution
do not need to be connected. However, if
\begin{equation}\label{eq:cond}
\sum_k k(k-2)p_k>0
\end{equation}
(which translates into $z\!>\!1$ in the case of ER graphs), one of the
connected components (called the giant connected component) will
gather a finite fraction of all links and vertices \cite{Newman2001}.
This is a phenomenon akin to percolation. In Ref.~\cite{Newman2001}
the size of the giant component and the size distribution of finite
components were calculated. The degree distribution in the giant
component $p^{(g)}_k$ was calculated in Ref.~\cite{BauerBernard}.
Here we generalize those results and calculate the two-point
distributions $p^{(g)}_{q,r}$ and $\kb^{(g)}(k)$ for the giant
component.

We will use the method of generating functions introduced in
\cite{Newman2001}. The crucial observation is that the finite
connected components are essentially trees. That is because a link
emerging from one of the vertices in the component has the
probability $\propto s/V$ of connecting back to a node from this
component, where $s$ is the size of the component. So for finite $s$
this becomes negligible in the large-$V$ limit.

Now let us pick a link from the graph at random. It belongs to some
connected component. We will call $P_1(s)$ the probability that
cutting this link will split the component into two parts, one of
them finite and having size $s$.  Stated differently, $P_1(s)$ is the
probability that a randomly chosen link will lead into a finite part
of size $s$.  By the argument above this finite ``half'' will be a
tree.  Because of that, one can write down the equation for the
generating function $H_1(x)=\sum_s P_1(s) x^s$ \cite{Newman2001}:
\begin{equation}\label{eq:h1gen}
H_1(x)=x G_1(H_1(x)),
\end{equation}
where
\begin{equation}\label{eq:g1gen}
G_1(x)=\frac{G'_0(x)}{G'_0(1)}=\frac{1}{z}G'_0(x),
\quad
G_0(x)=\sum_{k=0}^\infty p_k x^k.
\end{equation}
We denote by $u$ the value of $H_1(1)$:
 \begin{equation}\label{def:u}
u\equiv H_1(1)=\sum_s P_1(s).
\end{equation}
When there is no giant component in the graph, all connected
components are finite and are trees. This means that cutting each
link will result in two finite parts; thus, $u\!=\!1$ However, when the
giant component appears, then there is a nonzero probability
that the chosen link will belong to this component and either
cutting it will split the component into two infinite parts, or will
not split it at all. As this probability is missing from $P_1(s)$
the sum \eqref{def:u} will be smaller the one. $u$ is to be
interpreted as the probability that a randomly chosen link is
connected to a finite part on at least one side of the graph
\cite{fronczak}.  It follows that $u^2$ is the probability that a
random link belongs to a finite component of arbitrary size.

That can be derived in a more explicit way. Let us denote by
$P_{1,1}(s)$ the probability that a randomly chosen link belongs to
a component of size $s$. Then,
\begin{equation}
P_{1,1}(s)=\sum_{t=0}^s P_1(t)P_1(s-t).
\end{equation}
It is a \textit{convolution} of the probability distribution
$P_1(s)$ with itself, so its generating function is just $H^2_1(x)$.
Then $u^2=H^2_1(1)=\sum_s P_{1,1}(s)$ is the probability that a link
belongs to a finite connected component of arbitrary size and
$1\!-\!u^2$ is the probability that it is inside the giant component.

Finally, if we denote by $P_0(s)$ the probability that a randomly
chosen {\em vertex} belongs to a finite component of size $s$, we
can obtain its generating function $H_0(x)$ from
$H_1(x)$~\cite{Newman2001}:
\begin{equation}\label{eq:h0gen}
H_0(x)\equiv\sum_s P_0(s)x^s=x G_0(H_1(x)).
\end{equation}
By the same arguments as above,
\begin{equation}\label{eq:h}
h\equiv H_0(1)=G_0(u)
\end{equation}
is the probability that a randomly chosen vertex belongs to a finite
connected component and $1\!-\!h$ is the probability that it belongs to
the giant component.

It follows from \eqref{eq:h1gen} and \eqref{def:u} that $u$ is the
solution of the equation
\begin{equation}\label{eq:u}
u=G_1(u).
\end{equation}
From the definition \eqref{eq:g1gen} it is easy to note that $u\!=\!1$
is always a solution, but when condition \eqref{eq:cond} is
fulfilled the above equation has a solution smaller than 1 as well
\cite{Newman2001}. As argued, this signals the appearance of a giant
component.

\subsection{Average degree}

Using the results of the previous section it is easy to derive
formulas for the average degree in the giant component $\zg$ and in
the rest of the graph $\zf$:
\begin{subequations}
\begin{eqnarray}
\label{eq:zg2}
z^{(g)}&=&\av{\frac{2L^{(g)}}{V^{(g)}}}=z\frac{1-u^2}{1-h},\\
z^{(f)}&=&\av{\frac{2L^{(f)}}{V^{(f)}}}=z\frac{u^2}{h}.
\end{eqnarray}
\end{subequations}

As we have already pointed out, the giant connected component is
not a tree. The number of independent loops that it contains equals
\begin{equation}\label{eq:nloops}
L^{(g)}-V^{(g)}+1\approx V \left(\frac{z}{2}(1-u^2)-1+h\right),
\end{equation}
and as all the remaining connected components are trees, this is
also the number of loops in the whole graph.

We can also easily calculate the number of finite connected components
$n_{cn}$ knowing that they form a forest. The number of links in the
forest is $L^{(f)}=V^{(f)}-n_{cn}$ which gives
\begin{equation}
\av{n_{cn}}=\left(h-u^2\frac{z}{2}\right) V.
\end{equation}

From that we can derive the formula for the average size of the finite
connected component:
\begin{equation}
\av{s}^{(f)}=\av{\frac{V^{(f)}}{n_{cn}}}=\frac{2h}{2h -u^2 z}.
\end{equation}

\subsection{Degree distribution}

In this section we will calculate the degree distribution $\pf_k$ in
the nongiant component part of the graph. From the relation
\begin{equation}\label{eq:pg+pf}
p_k=(1-h)\pg_k+h \pf_k,
\end{equation}
we automatically get the distribution $\pg_k$ in the giant
component.  This has been already done in \cite{BauerBernard}, but we
find it instructive to use the same method of generating functions
as described in Sec.~\ref{sec:connected}.  The idea is to apply it
only to the graph with the giant component excluded---i.e., to the
finite connected components.  We will use a tilde to denote the
generating functions of the sought probability:
\begin{equation}\label{def:gtilde}
\widetilde G_0(x)=\sum_{k=0}^\infty \pf_k x^k, \quad \widetilde
G_1(x)=\frac{\widetilde G'_0(x)}{\widetilde G'_0(1)}.
\end{equation}
Using the argument from Ref.~\cite{Newman2001} we obtain the same
equations
\begin{subequations}
\label{eq:htilde}
\begin{eqnarray}
\label{eq:htildea}
\widetilde H_1(x) & = & x \widetilde G_1(\tilde H_1(x)),\\
\label{eq:htildeb}
\widetilde H_0(x) & = & x \widetilde G_0(\tilde
H_1(x)),
\end{eqnarray}
\end{subequations}
for the generating functions of the probabilities $\widetilde
P_1(s)$ and $\widetilde P_0(s)$. Here $\widetilde P_0(s)$ is the
probability that a vertex belongs to a finite component of size $s$
provided that it belongs to a finite component and $\widetilde
P_1(s)$ is the probability that a link leads into a finite component
of size $s$ provided that it leads into a finite component. From
this we can write the relations
\begin{equation}
P_0(s)=h\widetilde P_0(s),
\quad
P_1(s)=u\widetilde P_1(s),
\end{equation}
which leads to
\begin{subequations}
\label{eq:hht}
\begin{eqnarray}
\label{eq:hhta}
H_0(x)&=&h \widetilde H_0(x),\\
\label{eq:hhtb}
H_1(x)&=&u \widetilde H_1(x).
\end{eqnarray}
\end{subequations}

To solve Eqs.~\eqref{def:gtilde}, \eqref{eq:htilde}, and
\eqref{eq:hht} for $\pf_k$ we make an ansatz
\begin{equation}
p^{(f)}_k = \frac{p_k a^k }{G(a)}.
\end{equation}
Then,
\begin{align}
\widetilde G_0(x)=\frac{G_0(x a)}{G_0(a)}, \quad
\widetilde G_1(x)=\frac{G_1(x a)}{G_1(a)},
\end{align}
so that Eq.~\eqref{eq:htildea} can be rewritten as
\begin{equation}
a \widetilde  H_1(x)=\frac{a}{G_1(a)} x G_1(\widetilde H_1(x) a).
\end{equation}
Comparing with \eqref{eq:h1gen} we see that it will be fulfilled if
\begin{equation}
a \widetilde H_1(x)  = H_1\left(\frac{a}{G_1(a)}x\right).
\end{equation}
Inserting this into \eqref{eq:hhtb} we get
\begin{equation}
a  H_1(x)  = u H_1\left(\frac{a}{G_1(a)}x\right),
\end{equation}
because of Eq.~\eqref{eq:u}, which can be solved by putting $a\!=\!u$.

Now we must check Eq.~\eqref{eq:hhta}.  Using Eqs.~\eqref{eq:h0gen},
\eqref{eq:htildeb}, and \eqref{eq:hhtb} we get
\begin{equation}\begin{split}
    h \widetilde H_0(x)= &\; h x \tilde G_0(\widetilde H_1(x)) =h x\frac{G_0(u \widetilde H_1(x) )}{G_0(u)}\\
    = &\; x h \frac{G_0(H_1(x) )}{h}=H_0(x).
\end{split}
\end{equation}
So finally,
\begin{equation}\label{eq:pqf}
\pf_k=\frac{p_k u^k}{h}.
\end{equation}
From that and relation \eqref{eq:pg+pf} we get the formula for the
degree distribution in the giant component:
\begin{equation}\label{eq:pqg}
\pg_k=p_k \frac{(1-u^k)}{1-h}.
\end{equation}
In the limit $u\rightarrow1$ and $h\rightarrow1$ this reduces to
\begin{equation}\label{eq:pg-tree}
\pg_k=\frac{k}{z}p_k.
\end{equation}
In this limit the connected giant cluster is a tree.  Indeed, one
can check that
\begin{equation}\label{eq:two}
\sum_k k \pg_k = \sum_k \frac{k^2}{z}p_k = 2.
\end{equation}
To see this we must first note that Eq.~\eqref{eq:u} has always the
solution $u\!=\!1$. It becomes the only one when $G'_1(1)=1$, which is
equivalent to the condition \eqref{eq:two}.

\begin{figure}[!t]
\includegraphics[height=\figheight]{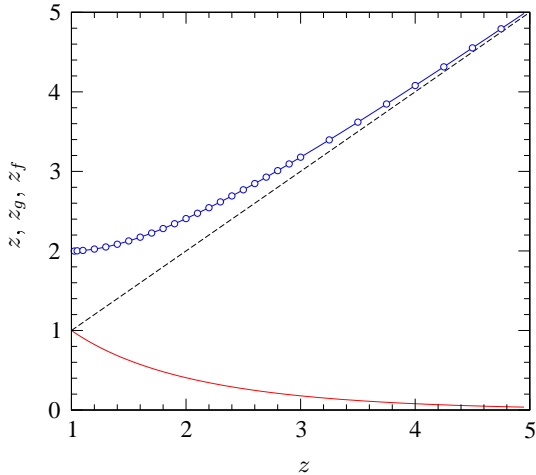}
\caption{\label{fig:zfg}Average degree $z$ (dashed
line), average degree $z_g$  of the connected component (upper solid
line), and average degree of the rest $z_f$ (lower solid line) as a
function of $z$ for ER graphs. Circles
  mark the results of MC simulations.}
\end{figure}

\subsection{Correlations}
\label{sec:correlations}

To calculate $p^{(g)}_{q,r}$ we use the relation
\begin{equation}
n_{q,r}(G)=n^{(g)}_{q,r}(G)+n^{(f)}_{q,r}(G).
\end{equation}
We have already assumed that vertex degrees are uncorrelated;  we
further assume that this is also true for the finite connected
components (nongiant) part of the graph. Assuming self-averaging
and using Eq.~\eqref{eq:indepnqr} for $n_{q,r}$ and $n^{(f)}_{q,r}$
we obtain
\begin{equation}\label{eq:pqr}
p^{(g)}_{q,r}=\frac{q  p_q r p_r}{z^2}
\frac{1}{1-u^2}\left(1-\frac{u^q u^r}{u^2}\right)
\end{equation}
and
\begin{equation}\label{eq:qqb}
\kb^{(g)}(k) = \frac{\av{k^2}}{z}\frac{1}{1-u^k}
\left(1-\frac{\av{k^2}^{(f)}}{z^{(f)}}\frac{z}{\av{k^2}}u^{k}\right).
\end{equation}
In the derivation we have used the relation
$\av{\frac{A}{B}}=\frac{\av{A}}{\av{B}}$, which should be valid for
self-averaging quantities in the large-$V$ limit.  Comparing this
with formulas \eqref{eq:indepnqr} and \eqref{eq:indepqbq} we note
that the correlations disappear in the limit $u \! \rightarrow \! 0$. In
the tree limit $u \!\rightarrow\! 1$ the formulas above take the form
\begin{equation}
\lim_{u,h\rightarrow 1} p^{(g)}_{q,r}=(q+r-2)\frac{1}{2}\frac{q  p_q
r p_r}{z^2}
\end{equation}
and
\begin{equation}
\lim_{u,h\rightarrow 1}\kb^{(g)}(k)=\frac{1}{z k}\left(
(k-2)\av{k^2} + \av{k^3}\right).
\end{equation}

\section{Examples}
\label{sec:examples}

\begin{figure}[!t]
\includegraphics[height=\figheight]{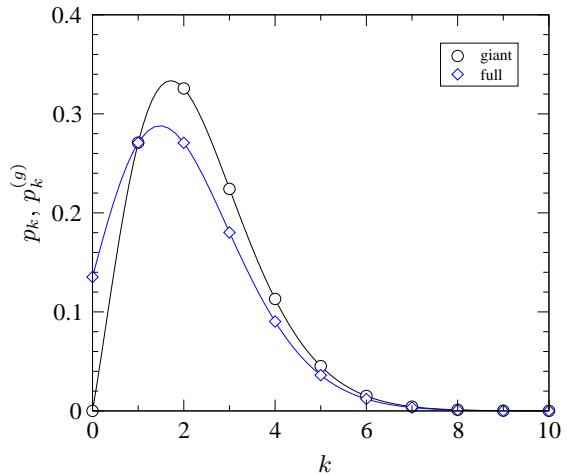}
\caption{\label{fig:pger}Degree distribution for ER
graphs with
  $z\!=\!2$.  Circles mark the results of MC simulations for the giant component and
  diamonds for the full graph. Solid lines denote analytical solutions.}
\end{figure}

While deriving our formulas we have made several assumptions: (i)
the vertex orders are uncorrelated, (ii) the measured quantities are
self-averaging, and of course (iii) all the derivations are only
valid in the large-$V$ limit. To check to what extent those
assumptions are satisfied and, more importantly, to check the
magnitude of the finite size effects, we have compared our
predictions to the results of MC simulations of moderate-sized
graphs (5000 vertices). To simulate ER graphs we used a
straightforward algorithm which connects vertices at random. To
generate maximally random graphs with a given distribution we used
the method described in
Refs.~\cite{BurdaKrzywicki,BogaczBurdaWaclaw} and implemented in
Ref.~\cite{graphgen}. This method consists of generating graphs with
suitably chosen one-point weights using a Metropolis-type
algorithm.

\subsection{Erd\"{o}s-R\'{e}nyi graphs}
For ER graphs the distribution $p_k$ is Poissonian,
\mbox{$p_k=e^{-z}\frac{z^k}{k!}$} and
\begin{equation}
G_0(x)=G_1(x)=e^{z(x-1)}.
\end{equation}
It follows that $H_1(x)=H_0(x)\equiv H(x)$, so $h\!=\!u$ with $h$ being
the closest to one (from below) positive solution of the equation
\begin{equation}\label{eq:her}
h= e^{ z (h-1)}.
\end{equation}
The results for $z^{(f)}$ and $z^{(g)}$ are shown in
Fig.~\ref{fig:zfg}.
They are compared with the results of the MC simulations of ER
graphs. The agreement is perfect, and there are no visible finite-size
effects (error bars are smaller than the size of the points).
The degree distribution can be now easily obtained from
\eqref{eq:pqg}. The results are presented in Fig.~\ref{fig:pger}.
Again, the agreement is very good without any noticeable finite-size
effects.

\begin{figure}[!t]
\includegraphics[height=\figheight]{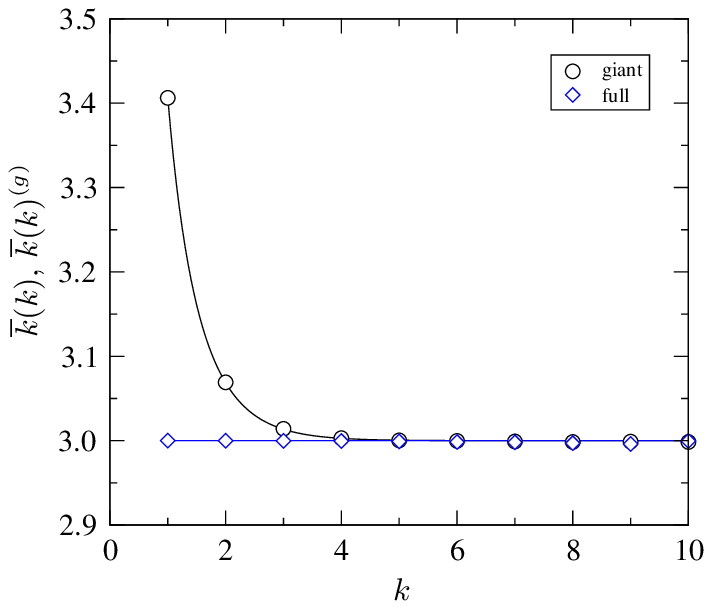}
\caption{\label{fig:qber}$\kb(k)$ for ER graphs with
  $z\!=\!2$.  Circles mark the results of MC simulations for the giant component and
  diamonds for the full graph; solid lines stand for analytical solutions.}
\end{figure}

In this case it may be instructive to derive those results in a
simpler way: when we omit the giant component from our
considerations we are left with a graph with $h N$ vertices and $h^2
L$ links on average. As there are no further restrictions,  we can
assume that this graph is an Erd\"{o}s-R\'{e}nyi graph as well.
This means that its degree distribution is again Poissonian with
mean $z^{(f)}$:
\begin{equation}
p^{(f)}_k = e^{-z_f}\frac{(z^{(f)})^k}{k!}= e^{-h z}\frac{z^k
h^k}{k!}.
\end{equation}
From the relation $h\!=\!u$ we obtain formula \eqref{eq:pqf}.
Finally, for $\kb(k)$ we get
\begin{equation}\label{eq:qber}
\kb^{(g)}(k)=\frac{z+1}{1-h^k}\left(1-\frac{z h+1}{z+1}h^k\right).
\end{equation}
The results are presented in Fig.~\ref{fig:qber}.
One can see clearly the appearance of correlations in the giant
connected component as advocated in the introduction. The agreement
with the predicted values is again very good.

\subsection{Exponential degree distribution}
\vspace{-1mm}
\begin{figure}[!t]
\includegraphics[height=\figheight]{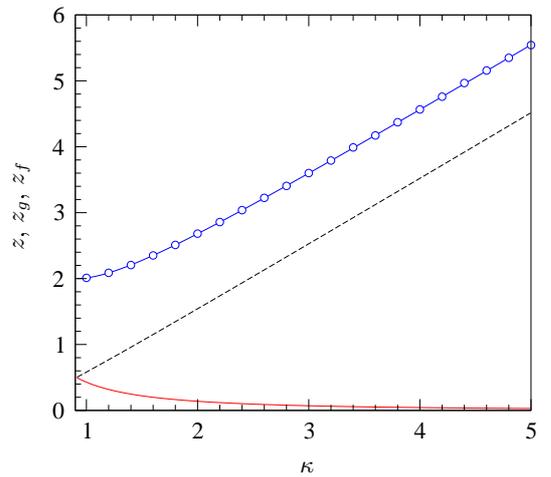}
\caption{\label{fig:zexp}Average degree $z$ (dashed
line), average degree $z_g$  of the connected component (upper solid
line), and average degree of the rest $z_f$ (lower solid line) as a
function of $\kappa$ for graphs  with exponential degree
distribution. Circles mark the results of MC simulations.}
\vspace{-4mm}
\end{figure}

As the second example we take graphs with exponential degree
distribution
\begin{equation}\label{eq:expdistribution}
p_k=(1-e^{-1/\kappa})e^{-k/\kappa}.
\end{equation}
The average degree in this case is
\begin{equation}
\vspace{-1.5mm}
  z=\frac{e^{-\frac{1}{\kappa}}}{1-e^{-\frac{1}{\kappa}}}
\approx\kappa-\frac{1}{2},\quad \kappa\gg 1,
\end{equation}
\vspace{-2mm}
and \cite{Newman2001}
\begin{equation}
G_0(x)=\frac{1-e^{-1/\kappa}}{1-x e^{-1/\kappa}},\quad
G_1(x)=G_0^2(x).
\end{equation}
This implies $u\!=\!h^2$. The giant component appears for $\kappa\!>\!1/\ln
3\approx 0.91$. The results for $z^{(g)}$ and $z^{(f)}$ are
presented in Fig.~\ref{fig:zexp}. As in the previous example, 
there are no visible deviations from the
theoretical predictions.

\begin{figure}[!t]
\includegraphics[height=\figheight]{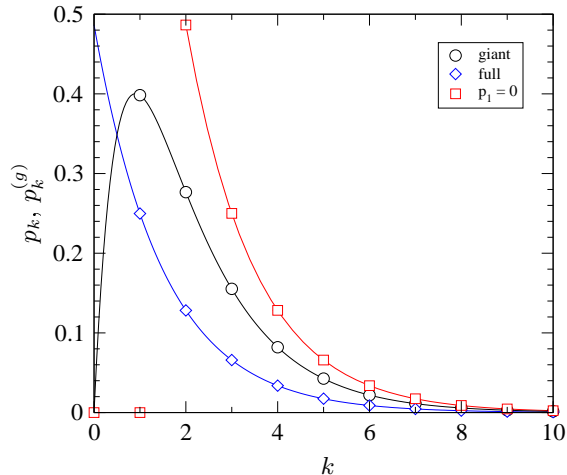}
\caption{\label{fig:pexp}Degree distribution for
graphs with exponential
  degree distribution with $\kappa\!=\!1.5$. Circles mark the results of MC simulations for the
  giant component and diamonds for the full graph; squares stand for the special case of connected graphs without leaves described in Sec.~\ref{sec:uncorrelatedconnectedgraphs}. Solid lines denote analytical solutions.}
\end{figure}

\begin{figure}[!t]
\includegraphics[height=\figheight]{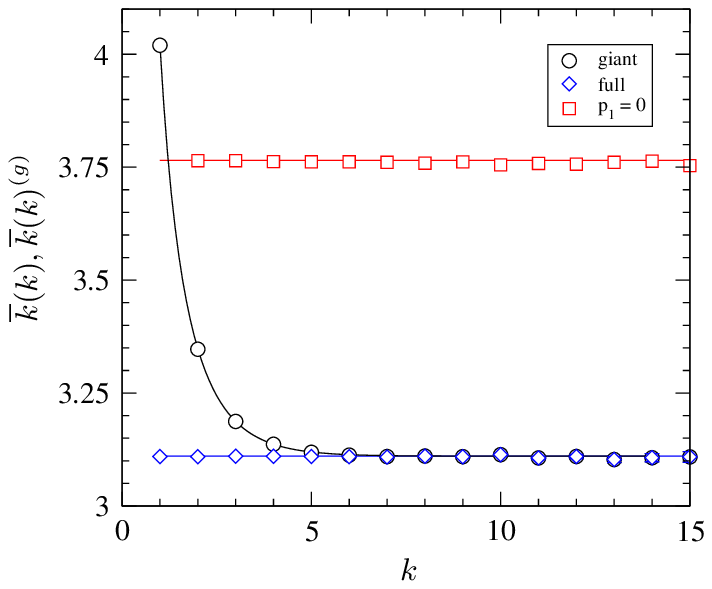}
\caption{\label{fig:qbexp}$\kb(k)$ for graphs with
exponential
  degree distribution  with $\kappa\!=\!1.5$. Circles mark the results of MC simulations for the
  giant component and diamonds for the full graph; squares stand for the special case of connected graphs without leaves described in Sec.~\ref{sec:uncorrelatedconnectedgraphs}. Solid lines denote analytical solutions.}
\end{figure}

In Figs.~\ref{fig:pexp} and \ref{fig:qbexp} results for $\pg_k$ and
$\kb^{(g)}(k)$ are presented for $\kappa\!=\!1.5$. We observe the same
kind of correlations in the giant component as in the case of ER
graphs.

\subsection{Scale-free graphs}
\label{sec:sf}


Probably the most interesting case are scale-free graphs with 
distribution $p_k \sim k^{-\beta}$.  While studying them we have to
consider two scenarios $2\!<\!\beta\!\le\!3$ and $\beta\!>\!3$.  In the
first case we expect correlations between node degrees, as pointed out in Refs.
\cite{BurdaKrzywicki,dogorovtsev,bpv,cbp}. This invalidates both the
derivation of Eqs.~\eqref{eq:h1gen} and \eqref{eq:pqr}. Additionally
the quantity $\av{k^2}$ diverges and so $\kb(k)$ is not
defined. Because our aim was to investigate the correlations appearing
solely as an effect of the connectedness of graphs, we have decided not
to study the $\beta\!\le\!3$ case in this paper.  This is, however, an
interesting issue and merits further investigation. One line of
pursuit is to use the algorithm proposed in \cite{cbp} to generate
uncorrelated graphs with heavy tails.  Then one should obtain
predictions at least for the joint probability $p_{q,r}$ which does
not contain any divergences.  One could also use  the $V$-dependent
``cutoff'' distribution as proposed in \cite{cbp} instead of the
``full'' distribution $p_k \sim k^{-\beta}$. This would yield
the $V$ depending results, but may not be feasible analytically. In the
case of $\beta\!<\!2$ already the first moment of the distribution $p_k$
is not defined and the generating function approach fails completely.

\begin{figure}[!t]
\includegraphics[height=\figheight]{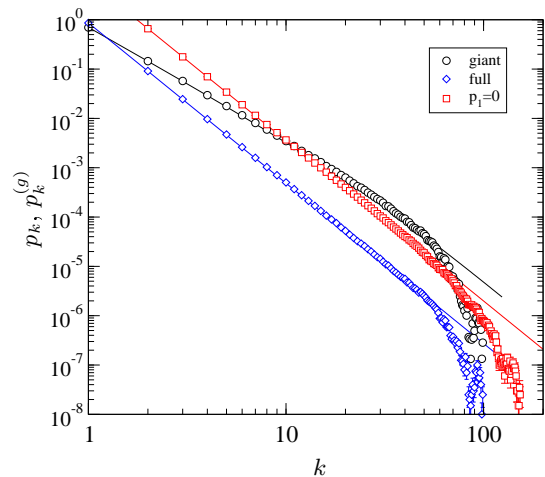}
\caption{\label{fig:sfpg}Degree distribution for
  scale-free graphs with~$\beta\!=\!3.25$. Circles mark the results for
  the giant component and diamonds for the full graph; squares stand
  for the special case of connected graphs without leaves described in
  Sec.~\ref{sec:uncorrelatedconnectedgraphs}. Solid lines denote
  analytical solutions.}
\end{figure}

\begin{figure}[!t]
\includegraphics[height=\figheight]{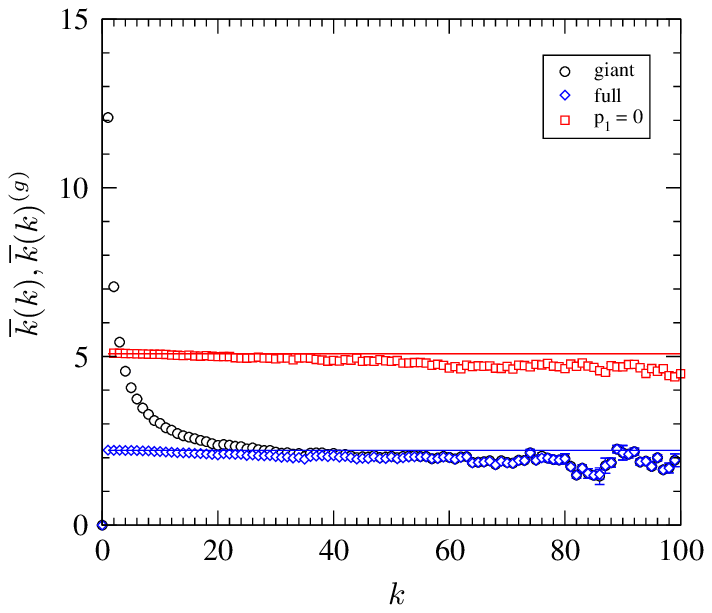}
\caption{\label{fig:sfqq}$\kb(k)$ for scale-free
  graphs with $\beta\!=\!3.25$. Circles mark the results for the giant
  component and diamonds for the full graph; squares stand for the
  special case of connected graphs without leaves described in
  Sec.~\ref{sec:uncorrelatedconnectedgraphs}. Solid lines denote
  analytical solutions.}
\end{figure}

When $\beta\!>\!3$ the $\av{k^2}$ is finite and there are no
correlations, at least in the infinite-size limit \cite{bpv,cbp}.
However, for finite $V$ we expect strong finite-size effects for
$\beta$ close to 3. To see this let us estimate the asymptotic
behavior of $\av{k^2}$:
\begin{equation}
\av{k^2}\approx \sum_k k^2 p_k -\int^\infty_{k_{c}(V)} k^2 p_k\approx
\av{k^2}_\infty - c V^{-\frac{\beta-3}{\beta-1}}.
\end{equation}
In the above we have assumed the natural cutoff {$k_c(V)\sim
V^{\frac{1}{\beta-1}}$} \cite{dogorovtsev,BurdaKrzywicki,bpv,cbp}. For
$\beta$ close to 3, this converges very slowly.  To observe
those effects we have simulated our system at $\beta\!=\!13/4$, when $\av{k^2}$ approaches its asymptotic value as $V^{-1/9}$.
The results of our simulations of graphs with 5000 vertices are
presented in Figs.~\ref{fig:sfpg}~and~\ref{fig:sfqq}. As expected the data for $p_k$ and $\pg_k$ distributions show
strong cutoff effects around $k\!=\!40$, but for smaller values of $k$
the agreement with theoretical predictions is rather good.
Looking at the results for $\kb(k)$ we notice two things: (i) Data for
the full graph show a deviation from a straight line, indicating the
presence of some correlations due to heavy tails. (ii) Data for the
giant connected component show a very strong effect of
correlations. The agreement with theoretical values is very poor, so
we have not included them in the picture. This is due to the described
cutoff effect on $\av{k^2}$. We can obtain a better agreement if we use in Eq. \eqref{eq:qqb} 
 the actual value of $\av{k^2}$ measured in simulations instead of its infinite-volume limit.

\section{Connected graphs}
\label{sec:connectedgraphs}

Finally, we would like to calculate the properties of the maximally
random connected graphs.  To this end we assume that the ensemble of
giant connected components of the maximal entropy graphs with
distribution $p_k$ is a maximal entropy ensemble of connected graphs
with distribution $\pg_k$ (we neglect the fluctuations in the number
of vertices and links of the giant component). This is a plausible
assumption as we do not put any additional constraints except
connectivity. In Appendix~\ref{app:entropy} we provide a more
detailed argumentation. With this assumption the properties of the
maximal entropy connected random graphs with distribution $\pg_k$
and/or average degree $z^{(g)}$ are the same as that of the maximal
entropy random graphs with distribution $p_k$ and/or average degree
$z$ given by Eqs.~\eqref{eq:pqg} and \eqref{eq:zg2}.

\subsection{Connected ER graphs}
\label{sec:con-er}

\begin{figure}
\includegraphics[height=\figheight]{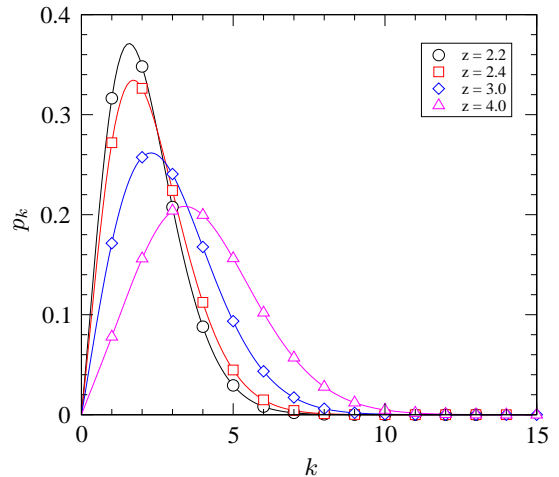}
\caption{\label{fig:pg}Degree distribution $p_k(k)$
in connected ER graphs with various average degrees. Points mark the
results of MC simulations, while solid lines denote analytical
solutions. The size of each graph is 5000 vertices.}
\end{figure}

\begin{figure}
\includegraphics[height=\figheight]{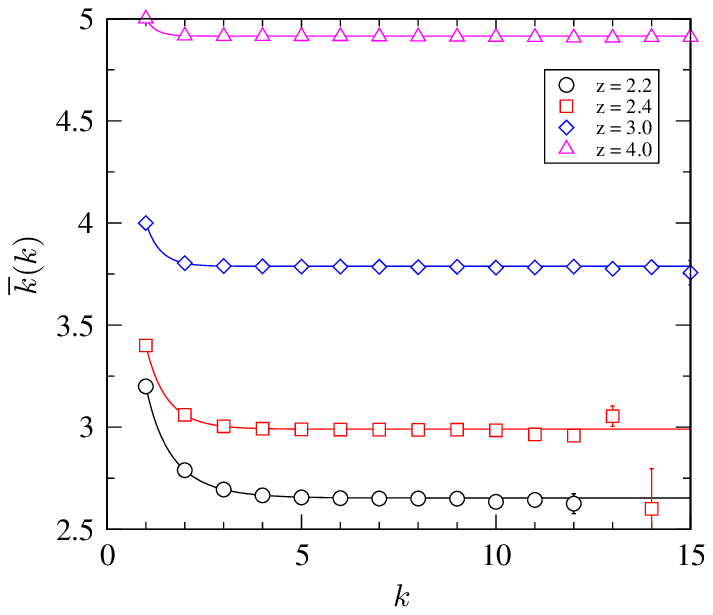}
\caption{\label{fig:qq}$\kb(k)$ for connected ER
graphs with various
  average degrees. Points mark the results of MC simulations, while solid lines denote analytical solutions. The size of each graph is 5000 vertices.}
\end{figure}

By connected ER graphs we mean maximal entropy connected graphs with
a given average degree $z^{(g)}$. According to the arguments from
the previous section this ensemble corresponds to the ensemble of
giant components in ER graphs with average degree $z$ related by
Eq.~\eqref{eq:zg2}.  For a given $z^{(g)}$ we solve this equation
for $z$ (numerically) and use formulas \eqref{eq:pqg} and
\eqref{eq:qber} for degree distribution and for $\kb(k)$
respectively. The results are presented in
Figs.~\ref{fig:pg}~and~\ref{fig:qq} and compared with the MC data
for connected graphs taken from \cite{oles}.
The agreement is very good which confirms the validity of the
assumption made in the previous section.


\subsection{Connected random graphs with arbitrary degree distribution}

To calculate the properties of connected random graphs with
arbitrary degree distribution we need to invert Eq.~\eqref{eq:pqg}.
This can be done by rewriting it as
\begin{equation}\label{eq:invert}
p_k=(1-h)\frac{\pg_k}{1-u^k},\quad p_0>0,
\end{equation}
where $u$ satisfies Eq.~\eqref{eq:u}:
\begin{equation}
u=\frac{\sum_{k=1}^\infty \pg_k  \frac{k u^{k-1}}{1-u^k}}
{\sum_{k=1}^\infty \pg_k  \frac{k}{1-u^k}}.
\end{equation}
The above equation can be solved by the simple iteration procedure. To
prove that it has a solution we rewrite it~as
\begin{equation}
\sum_{k=1}^\infty \pg_k k u\frac{1-u^{k-2}}{1-u^k}\equiv g(u) =0.
\end{equation}
It is easy to check that
\begin{equation}
g(0)=-\pg_1,\quad \lim_{u\rightarrow1}
g(u)=\sum_{k=1}^\infty \pg_k k - 2.
\end{equation}
So for connected graphs $g(1)$ is positive ($z^{(g)}\!\ge\!2$) and
$g(0)$ negative ($\pg_1\!\ge\!0$).

Once we know $u$ we can calculate $h$ and $p_0$ from the
normalization of the distribution $p_k$ and Eq.~\eqref{eq:h}:
\begin{equation}\begin{split}
1=p_0+(1\!-\!h)\sum_{k=1}^\infty\frac{\pg_k}{1\!-\!u^k},\quad
h=p_0+(1\!-\!h)\sum_{k=1}^\infty\frac{u^k \pg_k}{1\!-\!u^k}.
\end{split}
\end{equation}
Because
$\sum_{k=1}^\infty\frac{pg_k}{1-u^k}-\sum_{k=1}^\infty\frac{u^k
  pg_k}{1-u^k}=1$, those two equations are not independent and  we can
set $p_0\!=\!0$. Then,
\begin{equation}
h=1-\left(\sum_{k=1}^\infty\frac{\pg_k}{1-u^k}\right)^{-1}.
\end{equation}

\subsection{Simulating connected graphs}

This procedure may be actually used to generate connected random
graphs in an efficient way. Instead of generating connected graphs
with degree distribution $\pg_k$ and checking the connectivity after
every move, we can generate graphs with distribution $p_k$ given by
\eqref{eq:invert} and use the giant connected component. This still
requires calculating the connected parts, but it need to be done
only once before each measurement.

\begin{figure}
\includegraphics[height=\figheight]{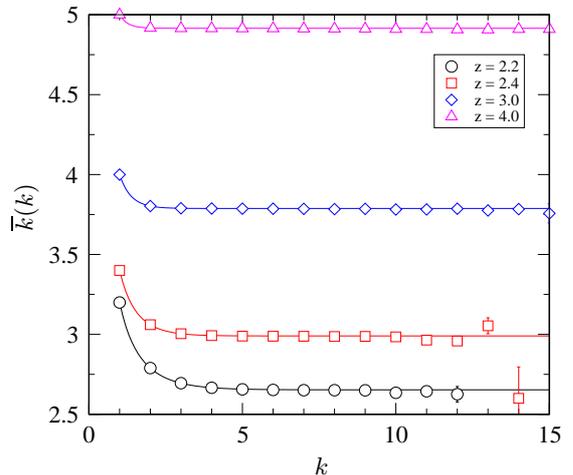}
\caption{\label{fig:inverse}Degree distribution $\pg_k$ in the
connected giant component. Circles mark the results of MC
simulation, while the solid line denotes the desired distribution
\eqref{eq:desire}. }
\end{figure}

As an example, we have generated connected maximally random graphs
with Poissonian  degree distribution
\begin{equation}\label{eq:desire}
p^{(g)}_k=e^{-z}\frac{z^k}{k!},\quad k>0,\quad p_0=0,
\end{equation}
with $z^{(g)}\!\approx\!2.7236$. For this distribution $u\!\approx\!
0.1209 $, $h\!\approx\!0.0341$, and $z\!\approx\!2.6696$. Using the program~\cite{graphgen} we have simulated a maximally random graph with
$5000/(1\!-\!h)\!\approx\!5177$ vertices and $6910$ links with degree
distribution \eqref{eq:invert}. We generated 10 000 independent
graphs. The average size of the giant component was $5000.24\pm0.25$
with standard deviation $\approx\! 20$. The degree distribution in the
connected component agrees very well with the desired one, as can be
seen in Fig.~\ref{fig:inverse}.

\section{Uncorrelated connected graphs}
\label{sec:uncorrelatedconnectedgraphs}

An interesting situation arises when $p_1\!=\!0$; i.e., vertices with
degree 1 (leaves) are forbidden. Then $u\!=\!0$ and $h\!=\!p_0$. This
means that the resulting graph consists of one giant connected
component and $p_0 V$ isolated vertices only. It is easy to
understand: finite connected components are trees, but there are no
trees without leaves, except the degenerated ones made of a single
vertex. If we additionally set $p_0\!=\!0$ then we will obtain a graph
containing only the giant component---i.e., a connected graph.

But as observed in Sec.~\ref{sec:correlations}, $u\!=\!0$ implies the
absence of correlations. That would support our argument made in the
Introduction about the role of the one-degree vertices in the
appearance of correlations in a connected graph. Using the results
of the previous section we can state that vertex degrees in the
maximal entropy random graphs are uncorrelated if and only if
$p_1\!=\!0$; i.e., there are no leaves in the graph.

As a check, we have carried out simulations with the exponential
degree distribution and no leaves:
\begin{equation}
p_k=\frac{1-e^{-1/\kappa}}{e^{-2/\kappa}} e^{-\frac{k}{\kappa}},
\quad k>1, \quad p_0=p_1=0,
\end{equation}
for $\kappa\!=\!1.5$ ($z\!\approx\!3.055$). The results for the giant
component which consisted on average of more the $99.9\%$ of the
whole graph are presented in Figs.~\ref{fig:pexp} and
~\ref{fig:qbexp} (squares). As predicted, vertices are uncorrelated
in the stark contrast to the $p_1\!>\!0$ case plotted in the same
figures.

We have also performed simulations for the scale-free distribution
$1/k^{13/4}$ and no leaves. The results are presented in
Figs.~\ref{fig:sfpg} and \ref{fig:sfqq} (squares). We see that
correlations are very much suppressed compared to the case when we
admit leaves (presented in the same figures). The slight remaining
correlation is due to long tails as explained in Sec.~\ref{sec:sf}.

\section{Summary}
\label{sec:summary}

In this paper we have studied the correlations in connected random
graphs.  We have extended the results of
Refs.~\cite{Newman2001,BauerBernard,fronczak} and calculated
correlations in the giant connected components of random graphs. We
argue that those correlations are related  to the presence of nodes
with degree 1, suggesting that the only cause of correlations is
the absence of ``hedgehogs.'' This has been already stated in
\cite{pb1} where it has been shown that in the grand-canonical
ensemble of arbitrary-sized trees, where ``hedgehogs'' appear,
correlations vanish. We find this to be a very interesting issue
that merits further studies.

The correlations observed in connected random graphs are an example
of the so-called ``structural'' or ``kinematic'' correlations, as
they appear in consequence of some global constraint. This should be
contrasted with ``dynamic'' correlations which are the result of
local two-point interactions between vertices. Such correlations
may be generated by two-point weights \cite{pb2}.  This distinction
can be important in simplicial quantum gravity where degree-degree
correlations are interpreted as curvature-curvature correlations
(see, for example, \cite{SmitBaker}). However, as the simplicial
manifolds are connected by definition those correlations are due to
the above described mechanism rather than to some kind of
gravitational interaction \cite{pb1,bbpt}. We believe that our
results may help in clarifying such issues and in the interpretation
of data obtained from MC simulations.

Finally, we have shown how to relate the giant connected components to
the maximal entropy connected graphs ensemble. This allowed us to
propose an efficient method for generating connected random graphs
based on the Metropolis algorithm.

\begin{acknowledgments}
  We would like to thank Zdzislaw Burda, Jerzy Jurkiewicz,
  Andrzej Krzywicki, and Bartłomiej Wacław for valuable discussions. This work
  was supported by KBN Grant No. 1P03B-04029 and EU Grants Nos.
  MTKD-CT-2004-517186 (COCOS) and MRNT-CT-2004-005616 (ENRAGE).
\end{acknowledgments}

\appendix

\section{Non-self-averaging ensemble}
\label{app:non-sa}

Denoting by $\mathcal{G}(V;k)$ the ensemble of all simple regular
graphs with $V$ vertices and degree $k$ (in a regular graph all
vertices have the same degree), we define
\begin{equation} \label{eq:nonsa}
\mathcal{G}(V)=\bigcup_k\mathcal{G}(V;k),\quad
P(G)=\frac{w_k}{\#\mathcal{G}(V;k)},
\end{equation}
where $\#\mathcal{G}(V;k)$ denotes the number of graphs in the
ensemble $\mathcal{G}(V;k)$ and $w_k$ is an arbitrary probability
distribution. With this definition we find
\begin{equation}\begin{split}
p_q &=\sum_{G\in\mathcal{G}}\frac{n_q}{V}P(G)=\sum_k\!\sum_{G\in\mathcal{G}(V;k)}\!\frac{w_k\delta_{k,q}}{\#\mathcal{G}(V;k)}\\
&=\sum_k w_k \delta_{k,q} =w_q.
\end{split}
\end{equation}
It is  easy to note that this poorly describes the distributions of
single graphs which are just $\delta$'s. The variance of $p_k$ is
\begin{equation}\begin{split}
\delta^2 p_q &=\sum_{G\in\mathcal{G}}(\frac{n_q}{V}\!-\!w_q)^2 P(G)
=\sum_k\!\sum_{G\in\mathcal{G}(V;k)}\!\!\!\frac{w_k(\delta_{k,q}\!-\!w_q)^2}{\#\mathcal{G}(V;k)}\\
&=\sum_k w_k (\delta_{k,q}\!-\!w_q)^2 =w_q - 2 w_q^2 + w_q^2 \sum_k w_k.
\end{split}
\end{equation}
and indeed does not disappear in the large-$V$ limit.

For correlations we obtain
\begin{equation}
\av{\frac{\tn qr}{2L}} =q r\av{\frac{n_q}{2L} \frac{n_r}{2L}}=\sum_k
w_k \delta_{q,k}\delta_{r,k}=w_q \delta_{q,r}
\end{equation}
and
\begin{equation}
q r\av{\frac{n_q}{2L}}\av{\frac{n_r}{2L}}=
\sum_{k}w_{k}\delta_{k,q}\sum_{k'}w_{k'}\delta_{k',r}=w_{q}w_{r}.
\end{equation}
So the condition \eqref{eq:indepnqr} is not satisfied. It means that
vertices on each particular graph are uncorrelated, but correlated
if the whole ensemble is considered.  This is easy to explain: if we
pick a link from a graph with a given $k$, then the information about
the first vertex does not provide any additional information;
however, if we do not know $k$, then the degree of the first vertex
will give us immediately the value of
its neighbor.\\

\section{Entropy of the giant connected components}
\label{app:entropy}

Let $\mathcal{G}$ and $P(G)$ define a maximal entropy ensemble with
$V$ vertices, $L$ links, and vertex degree distribution $p_k$. We
assume that the probability $P(G)$ factorizes:
\begin{equation}\label{eq:factor}
P(G)=\prod_{C\in G} P_c(C),
\end{equation}
where $C$ are the connected components of the graph $G$.

Let $\mathcal{G}_c$ denote the ensemble of all giant connected
components.  We assume that we can neglect the fluctuations, so all
the graphs in this ensemble have $V^{(g)}$ vertices and $L^{(g)}$
links. The degree distribution in this ensemble is $\pg_k$. Because
of the property \eqref{eq:factor}, the entropy \eqref{def:entropy}
of the whole ensemble $(\mathcal{G},P)$ is the sum of the entropy of
the giant connected component ensemble and the rest:
\begin{equation}\label{eq:entropy}
S=S^{(g)}+S^{(f)}.
\end{equation}
Now we assume that there exists a probability $P'_c$ defined on the
ensemble $\mathcal{G}_c$ such that the entropy
\begin{equation}
-\sum_{G\in\mathcal{G}_c}P'_c(G)\ln P'_c(G)
\end{equation}
is greater than $S^{(g)}$, but the vertex degree probability distribution
remains unchanged. Then we can define a new probability on the
ensemble $\mathcal{G}$:
\begin{equation}
P'(G)=P'_c(C^{(g)}) \prod_{C\neq C^{(g)}} P_c(C),
\end{equation}
where $C^{(g)}$ is the giant connected component of graph $G$.  The
degree distribution of the ensemble $(\mathcal{G},P')$ would be the
same as that of $(\mathcal{G},P)$ ensemble, but according to
\eqref{eq:entropy}, its entropy would be greater. This contradicts
the assumption that $(\mathcal{G},P)$ is the maximal entropy
ensemble and proves that the ensemble of giant connected components
is a maximal entropy ensemble.

\end{document}